# A Novel Thiophene-fused Ending Group Enabling an Excellent Small Molecule Acceptor for High-performance Fullerene-free Polymer Solar Cells with 11.8% Efficiency

*Dongjun Xie[†], Tao Liu[†], Wei Gao[†], Cheng Zhong, Lijun Huo, Zhenghui Luo, Kailong Wu, Wentao Xiong, Feng Liu,\* Yanming Sun\* and Chuluo Yang\**

D. Xie, W. Gao, Dr. C. Zhong, Z. Luo, K. Wu, Prof. C. Yang
Hubei Key Lab on Organic and Polymeric Optoelectronic Materials, Department of Chemistry, Wuhan University, Wuhan, 430072, P. R. China
E-mail: clyang@whu.edu.cn

T. Liu, Prof. L. Huo, W. Xiong, Prof. Y. Sun
Heeger Beijing Research and Development Center, School of Chemistry and Environment, Beihang University, Beijing 100191, P. R. China
E-mail: sunym@buaa.edu.cn

Prof. F. Liu
Department of Physics and Astronomy, and Collaborative Innovation Center of IFSA (CICIFSA), Shanghai Jiaotong University, Shanghai 200240, P. R. China
E-mail: fengliu82@sjtu.edu.cn

[†]These authors contributed equally to this work.

Fullerene-free polymer solar cells (PSCs) have attracted much attention since the breakthrough of power conversion efficiencies (PCEs) by using small molecular acceptors (SMAs).[1–11] Due to the well-tailored structure, SMAs can address fundamental issues that the traditional fullerene-based acceptors suffer from, such as poor absorption in visible region, fixed energy levels and high production cost.[12] Taking these progressive advantages into account, the SMAs show a promising future of replacing fullerene-based acceptors to develop high-performance solar cells.

Considerable efforts have been devoted to develop SMAs in broadening the absorption bands,[13–15] manipulating the highest occupied molecular orbital (HOMO) and the lowest unoccupied molecular orbital (LUMO) energy levels[3,16,17] and controlling crystallinity of SMAs.[8,16,18] Different electron-withdrawing groups (EGs), such as perylene diimide (PDI),[15,19–26] naphthalene diimide (NDI),[27] tetraazabenzodifluoroanthene diimide,[12,28] diketopyrrolopyrrole (DPP),[29] rhodamine, [30–32] 2-(3-oxo-2,3-dihydro-1*H*-inden-1-ylidene)malononitrile (INCN),[3,8,13,17,33–38] have been involved in the design of SMAs. Among these SMAs, the most successful one is ITIC (3,9-bis(2-methylene-(3-(1,1-dicyanomethylene)-indanone))-5,5,11,11-tetrakis(4-hexylpheny) -dithieno[2,3-*d*:2',3'-*d'*]-s-indaceno[1,2-*b*:5,6-*b'*]dithiophene), which was first reported by Zhan and coworkers.[38] ITIC shows a broad absorption spectrum (main absorption in the range of 500 -780 nm), appropriate HOMO and LUMO energy levels, high electron mobility, good crystallinity and film-forming ability. All of these features make ITIC an excellent model molecule to design new SMAs by systematically engineering the electron-donating core and the end-capping group. Several reports focusing on the modification of ITIC on electron-donating core IDTT (6,6,12,12-tetrakis(4-hexylphenyl)-6,12-dihydro-dithieno[2,3-*d*:2',3'-*d'*]-s-indaceno[1,2-*b*:5,6-*b'*] dithiophene) have been reported.[8,13,16,17] For example, Zhan and coworkers reported that the replacement of the side chains of IDTT from 4-hexylphenyl to 4-hexylthienyl downshifts molecular energy levels and enhances the intermolecular interactions;[16] Hou and coworkers reported that introducing electron rich π bridge expands the absorption band and further enhances the short circuit current ($J_{sc}$);[13] Li and coworkers reported that a side chain isomerism engineering on the alkylphenyl substituents of ITIC results in higher crystallinity and greater electron motility than those of ITIC.[8]

Compared to the abovementioned modification of ITIC by changing the central fused aromatic-ring-based core, very limited example was done to explore the influence of end-capping group on the properties of ITIC analogues. Hou and coworkers made a small modification for ITIC by introducing one or two methyl to the benzene of INCN (the ending group for ITIC), the resulting PCE amazingly exceeded 12% when using PBDB-T (poly[(2,6-(4,8-bis(5-(2-ethylhexyl)thiophen-2-yl)-benzo[1,2-*b*:4,5-*b*']dithiophene))-alt-(5,5-(1',3'-di-2-thienyl-5',7'-bis(2-ethylhexyl)benzo[1',2'-*c*:4',5'-*c*']dithiophene-4,8-dione))]) as donor of active layer.[3]

In fact, designing new ITIC analogues by altering the ending group is a challenging task due to the lack of suitable electron-deficient active methylene precursors and synthetic difficulties, which made this route hard to explore. In this work, we designed and synthesized a new active methylene compound, CPTCN (2-(6-oxo-5,6-dihydro-4*H*-cyclopenta[*c*]thiophen-4-ylidene)malononitrile), by replacing benzene of INCN with thiophene. Using CPTCN as the precursor, we synthesized a new ITIC analogue, ITCPTC (3,9-bis(2-methylene-(3-(1,1-dicyanomethylene)-cyclopentane-1,3-dione-[*c*]thiophen))-5,5,11,11-tetrakis(4-hexylphenyl)-dithieno[2,3-*d*:2',3'-*d*']-s-indaceno[1,2-*b*:5,6-*b*']dithiophene) (**Figure 1a**). Benefited from the better electron delocalization of CPTCN moiety and stronger intermolecular interactions induced by sulfur−sulfur interaction, [39]

ITCPTC exhibits a more planar structure, a red-shift absorption spectrum and a larger crystal coherence length in film when compared to ITIC. By using ITCPTC as acceptor and our recently reported wide- bandgap polymer PBT1-EH as donor (**Figure 1a**),[40] the resulting PSCs achieved impressive high PCEs of 11.8% with a remarkably high fill factor (FF) of 0.75, a nearly 20% improvement in PCE comparing to ITIC-based control device. To the best of our knowledge, these values are among the highest PCEs and FFs in fullerene-free PSCs.

The targeted compound of ITCPTC was synthesized through Knoevenagel condensation reaction between the active methylene compound of CPTCN and IT-CHO with a high yield of over 90% (Scheme S1). The new compound shows good solubility in common organic solvents, such as chloroform (CF), chlorobenzene (CB) and *o*-dichlorobenzene (*o*-DCB). The chemical structures of CPTCN and ITCPTC were characterized by $^1$H NMR, $^{13}$C NMR and high resolution mass spectrometry (HRMS). The thermogravimetric analysis (TGA) of ITCPTC indicates its good thermal stability with the decomposition temperatures ($T_d$) of 245 °C with 5% weight loss.

As shown in **Figure 1b**, the new compound of ITCPTC in dilute chloroform (CF) solution ($10^{-6}$ M) exhibited strong and broad absorption in the region of 500–740 nm with an maximum molar absorption coefficient of $2.9 \times 10^5$ M$^{-1}$ cm$^{-1}$ at 678 nm, which is larger than that of ITIC ($2.2 \times 10^5$ M$^{-1}$ cm$^{-1}$ at 665 nm). The maximum absorption peak of ITCPTC showed a near 13 nm red shifts compared with that of ITIC. In thin film (**Figure 1c**), the absorption spectra of ITCPTC and ITIC exhibited significant red shifts with respect to their corresponding solution spectra, with the absorption edge of ITCPTC exceeding that of ITIC by 10 nm. The red-shifted absorption spectra as well as the intensive shoulder peak in neat film indicate the strong π-π stacking interaction between molecules. In addition, the absorption spectra of ITCPTC and PBT1-EH are well complimented to ensure a full harvest of incident light in visible region by active layers (**Figure 1c**).

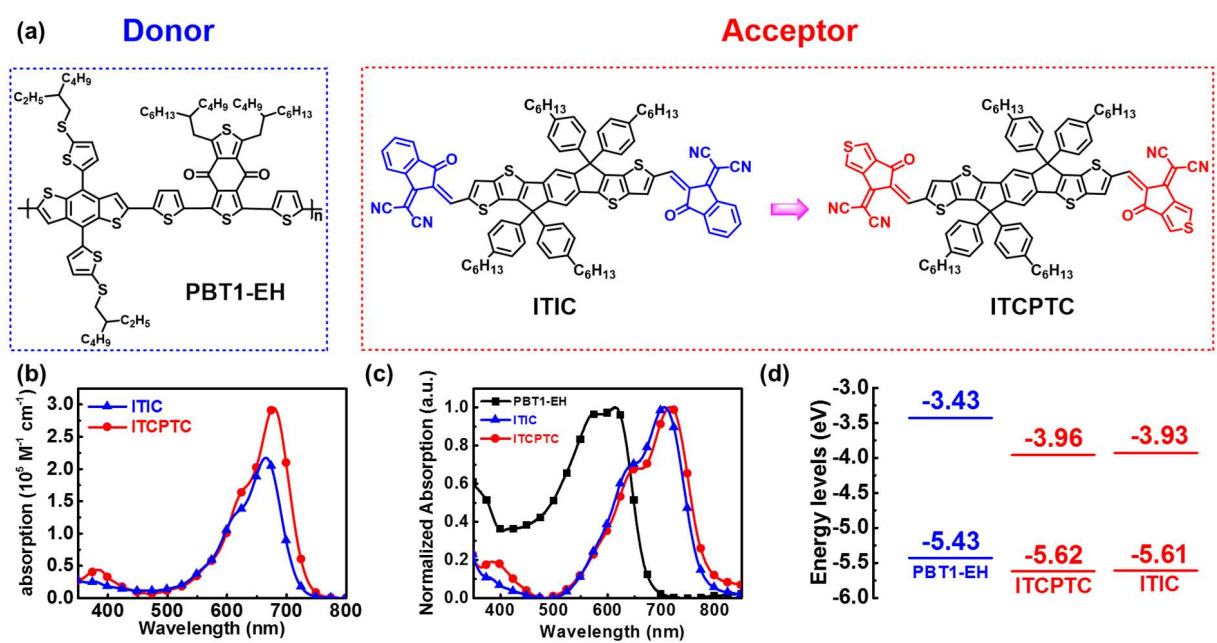

**Figure 1.** a) The chemical structures of donor and acceptor in this study. b) UV-vis absorption spectra of ITIC and ITCPTC in chloroform solution ($10^{-6}$ M). c) UV-vis absorption spectra of PBT1-EH, ITIC and ITCPTC in neat film. d) The energy level alignments of PBT1-EH, ITIC and ITCPTC.

The HOMO and LUMO energy levels of ITCPTC and ITIC were measured by cyclic voltammetry (CV) under the identical condition (Figure S1). The onset oxidation and reduction potentials of ITCPTC versus Fc/Fc$^+$ (0.44 V) were found to be 1.26 V and -0.40 V, respectively. The corresponding HOMO and LUMO energy levels of ITCPTC were determined to be -5.62 eV and -3.96 eV, respectively, which are slightly lower than those of ITIC (HOMO/LUMO: -5.61/-3.93 eV). The experimental values agree well with the calculated HOMO/LUMO levels of ITCPTC (-5.54/-3.45 eV) and ITIC (-5.49/-3.38 eV) (Figure S2). The lower LUMO energy level of ITCPTC over ITIC will have a negative effect on the open circuit voltage ($V_{oc}$) of ITCPTC-based PSCs according to the classical principle that $V_{oc}$ value is directly proportional to energy differences between the HOMO of electron donor and the LUMO of electron acceptor.[41] However, this can be compensated by choosing suitable polymer donors with deep HOMO energy level. In addition, ITCPTC shows more planar structure than ITIC with a small dihedral angle of 1.24 degree between CPTCN and IDTT units (Figure S2), which may be conducive to enhance intermolecular stacking and to improve the carrier mobility.

**Table 1.** The basic photophysical and electrochemical data of ITCPTC and ITIC.

| Acceptor | $\lambda_{max}$[a] (nm) | $\varepsilon_{max}$[a] ($M^{-1}$ $cm^{-1}$) | $\lambda_{onset}$[a] (nm) | $\lambda_{max}$[b] (nm) | $\lambda_{onset}$[b] (nm) | $E_g^{opt}$[c] (eV) | HOMO[d] (eV) | LUMO[d] (eV) | $E_g^{cv}$[e] (eV) | $\mu_e$[f] ($cm^2$ $V^{-1}$ $s^{-1}$) |
|---|---|---|---|---|---|---|---|---|---|---|
| ITIC | 665 | 2.9× 10$^5$ | 725 | 705 | 775 | 1.60 | -5.61 | -3.93 | 1.68 | 3.0×10$^{-3}$ |
| ITCPTC | 678 | 2.2× 10$^5$ | 740 | 720 | 785 | 1.58 | -5.62 | -3.96 | 1.66 | 3.2×10$^{-3}$ |

[a]In chloroform solution. [b]In neat film drop-cast from chloroform solution. [c]Calculated from $E_g^{opt} =1240/\lambda_{onset}$. [d]Obtained from cyclic voltammetry (CV) method. [e]$E_g^{cv} = E_{LUMO}-E_{HOMO}$. [f]Measured by space-charge-limited current (SCLC) method.

The electron mobilities of ITCPTC and ITIC were investigated by employing space-charge-limited current (SCLC) method with a device structure of ITO/Al/ITCPTC or ITIC/Al/ITO. The new compound of ITCPTC showed a high electron mobility of 3.2 × 10$^{-3}$ $cm^2$ $V^{-1}$ $s^{-1}$, which is in the same order of magnitude to that of ITIC (3.0 × 10$^{-3}$ $cm^2$ $V^{-1}$ $s^{-1}$). The high electron mobility of ITCPTC will benefit the balanced carrier transport and reduce the probability of carrier recombination on the way to electrodes when acted as electron acceptor. Grazing incident X-ray diffraction (GIXD) was used to investigate the structure order of ITCPTC and ITIC neat films to gain insight into the crystalline differences between two SMAs, and the corresponding line-cuts and 2D GIXD patterns are shown in **Figure 2**. Both ITIC and ITCPTC showed a (100) diffraction peak in in-plane direction and a π-π stacking (010) peak in out-of-plane direction, suggesting the dominant face-on crystal orientation. The thiophene replacement of benzene ring in INCN does not change the crystalline behavior and solid-state packing of molecules. Both molecules show (100) peaks location at 0.345 $Å^{-1}$, corresponding to a distance of 1.82 nm. The π-π stacking peaks for ITIC and ITCPTC are all at 1.79 $Å^{-1}$, corresponding to a distance of 0.35 nm. It should be noted that ITCPTC has more amorphous content in thin film, as seen from the 1.3 $Å^{-1}$ liquid scattering halo. Yet its crystal coherence length in π-π stacking direction is slightly larger (1.8 nm) comparing to ITIC (1.5 nm). Thus the thiophene modification of ITIC molecule leads to better persistence along π-π stacking direction normal to basal plane. And when hopping model is applied to investigate carrier transport, the high content amorphous region of ITCPTC in the percolation pathways does not slow down the carrier drifting, which results in SCLC measurement of ITCPTC neat film being slightly larger over ITIC. Thus we suspect ITCPTC molecule should have better mobility, as a result of the electronic structure modification of using thiophene-fused end groups, most likely due to the added S-S interactions. A larger surface roughness root mean square (RMS) of ITCPTC (0.665 nm) over that of ITIC (0.521 nm) was found (Figure S3). These values are quite close, and thus the SCLC mobility measurements are free of roughness or thickness variation, which made the thickness estimation in Mott-Gurney appropriate.

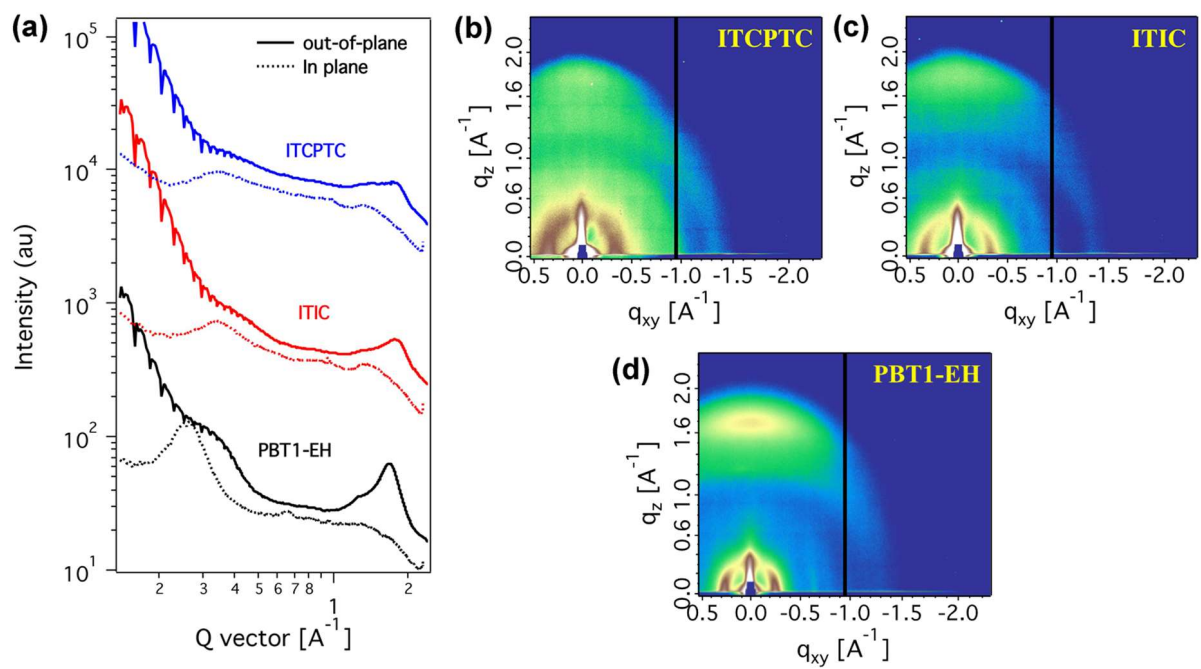

**Figure 2.** a) The in-plane and out-of-plane line-cut profiles of ITCPTC, ITIC and PBT1-EH neat films. 2D GIXD patterns: b) ITCPTC neat film; c) ITIC neat film; d) PBT1-EH neat films.

**Table 2.** Photovoltaic parameters of fullerene-free PSCs based on PBT1-EH:ITCPTC (1:1, w/w, 100 °C) and PBT1-EH:ITIC (1:1, w/w, 100 °C) in conventional structure under illumination of AM 1.5 G at 100 mW cm$^{-2}$.

| Active layer | $V_{oc}$ (V) | $J_{sc}$ (mA cm$^{-2}$) | FF | PCE$_{max}$ (%) | PCE$_{avg}$[a] (%) |
|---|---|---|---|---|---|
| PBT1-EH:ITCPTC | 0.95 | 16.5 | 0.751 | 11.8 | 11.4 ± 0.2 |
| PBT1-EH:ITIC | 0.99 | 15.7 | 0.634 | 9.8 | 9.7 ± 0.1 |

[a]The average PCEs were obtained from 20 devices.

In order to evaluate the new SMA of ITCPTC, PSCs have been fabricated with a device configuration of ITO/PEDOT:PSS/PBT1-EH:ITCPTC or ITIC /Zracac/Al (**Figure 3**a), where ITO (indium tin oxide) was used as the anode; PEDOT:PSS

(poly(3,4-ethylenedioxythiophene):poly(styrenesulfonate) ) served as the hole transporting layer (HTL); Zracac (zirconium acetylacetonate) acted as the cathode interlayer to lower the function of Al cathode;[42,43] The active layer consisted of donor PBT1-EH and SMAs (ITCPTC or ITIC). The control devices with PBT1-EH:ITIC active layer were fabricated under the same processing condition. The devices were optimized by screening the weight ratios of donor:acceptor and annealing temperatures, which have great effects on the short circuit current ($J_{sc}$) and fill factor (FF) (Table S1). The optimal weight ratio and annealing temperature were finally determined to be 1:1 and 100 ºC, respectively. The characteristic current density-voltage (*J-V*) curves and external quantum efficiency (EQE) spectra were shown in **Figure 3b-c**, and the key photovoltaic parameters were summarized in **Table 2**. The PBT1-EH:ITCPTC based device achieved a maximum PCE of 11.8% with a $V_{oc}$ of 0.95 V, a $J_{sc}$ of 16.5 mA cm$^{-2}$ and a FF of 0.751. The PBT1-EH:ITIC based control device acquired a $V_{oc}$ of 0.99 V, a $J_{sc}$ of 15.7 mA cm$^{-2}$, a FF of 0.634, yielding a PCE of 9.8%. Compared with the control device, though the ITCPTC based device showed a slightly declined $V_{oc}$, it revealed an enhanced $J_{sc}$ and a significantly increased FF, consequently a nearly 20% promotion in PCEs was achieved. The declined $V_{oc}$ for PBT1-EH:ITCPTC based device is consistent with the lower LUMO level of ITCPTC over ITIC. The enhanced $J_{sc}$ for PBT1-EH:ITCPTC based device can be partially attributed to a 10 nm red shift of the EQE spectrum of PBT1-EH:ITCPTC over PBT1-EH:ITIC (**Figure 3c**). The measured $J_{sc}$s of two PSCs were well matched with the corresponding integral values of EQE spectra within an error of 3%.

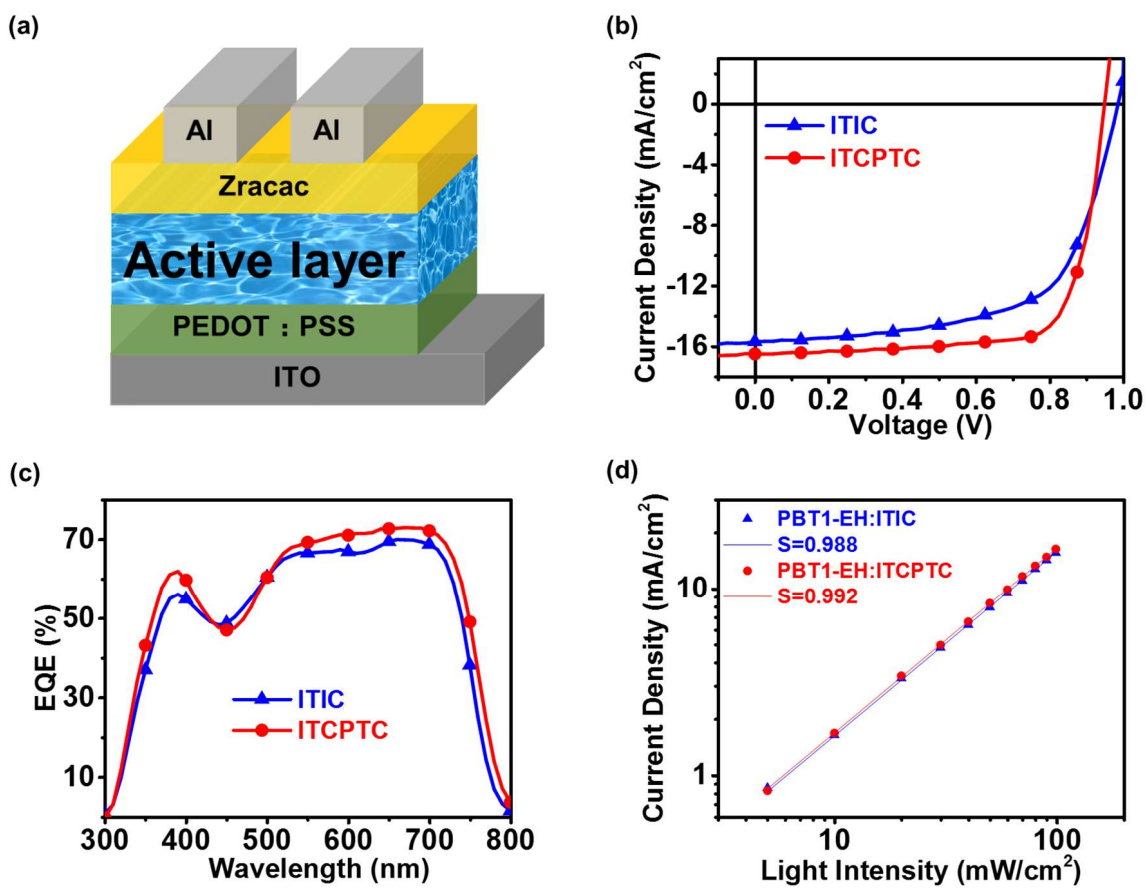

**Figure 3.** a) Device configuration of the studied PSCs. b) *J-V* characteristics curves and c) EQE spectra of the best performing PSCs of PBT1-EH:ITCPTC (1:1 w/w) and PBT1-EH:ITIC (1:1 w/w) annealed at 100°C under the illumination of AM 1.5 G at 100 mW cm$^{-2}$. d) $J_{sc}$ versus light intensity of PB1-EH:ITCPTC and PBT1-EH:ITIC in as-cast film.

Carrier mobility is an important factor influencing $J_{sc}$ and FF, the electron and hole mobility of two as-cast blend films were measured by using SCLC method for electron-only devices with the structure of ITO/Al/PBT1-EH:acceptor/Al/ITO and hole-only devices with the structure of ITO/MoO$_x$/PBT1-EH:acceptor/MoO$_x$/Al. The electron and hole mobility of PBT1-EH:ITCPTC and PBT1-EH:ITIC blend films were 2.69 × 10$^{-3}$ cm$^2$ V$^{-1}$ s$^{-1}$/1.71 × 10$^{-3}$ cm$^2$ V$^{-1}$ s$^{-1}$ and 2.59 × 10$^{-3}$ cm$^2$ V$^{-1}$ s$^{-1}$/1.52 × 10$^{-3}$ cm$^2$ V$^{-1}$ s$^{-1}$, respectively. The PBT1-EH:ITCPTC blend films showed slightly enhanced electron mobility and hole mobility compared to PBT1-EH:ITIC, and a fairly balanced carrier transport ($\mu_e/\mu_h$ = 1.57) which is favorable for charge extraction, and therefore benefitted to high $J_{sc}$ and FF.

Charge recombination occurred in the bulk heterojunction (BHJ) layer will reduce the amount of photogenerated carriers, leading to low PCEs. In order to obtain a better understanding of charge recombination in PSCs, the current densities ($J_{sc}$) under different light intensities were measured. The linear relationship between $J_{sc}$ and the light intensity in logarithmic coordinates ($\ln(J_{sc}) \propto \ln(P)$) were depicted in **Figure 3d**. In principle, in the short-circuit condition, $J_{sc}$ is in positive correlation with $P^S$, where S is a constant in a selected system reflecting the bimolecular recombination degree. If all dissociative charges can be collected at the electrodes before recombination, the S value could reach up to 1.[44–46] In other word, the larger the S, the smaller charge recombination occurs. As calculated, the slope (S) of $J_{sc}$ versus light intensity for PBT1-EH:ITCPTC is 0.992, which is larger than that of PBT1-EH:ITIC (0.988). The better charge extraction and less bimolecular recombination in PBT1-EH:ITCPTC active layer will reasonably enhance the utilization rate of excitons and improve the transport properties of carriers, facilitating the enhancement of $J_{sc}$ and FF.

To better understand the annealing role in the device performance, the surface morphology was investigated by using atomic force microscopy (AFM), transmission electron microscope (TEM), GIXD and resonant soft X-ray scattering (RSoXS). The as-cast blend films of PBT1-EH:ITCPTC and PBT1-EH:ITIC displayed smooth and uniform morphologies with root-mean-square (RMS) surface roughness of 0.85 and 1.06 nm (**Figure 4a and 4c**), respectively, thus no strong molecule segregation between PBT1-EH and two SMAs. When annealed at 100°C, the surface roughness of PBT1-EH:ITCPTC blend film was increased to 1.03 nm **(Figure 4b)**, while the value of PBT1-EH:ITIC blend film kept almost unchanged (1.07 nm) **(Figure 4d)**, suggesting that the annealed PBT1-EH:ITCPTC films underwent stronger phase separation. With the help of annealing, the blend film of PBT1-EH:ITCPTC exhibited clear nano-fibrillar structures as shown in AFM and TEM images (Figure S4), which is beneficial for charge transport. The 2D GIXD patterns and the corresponding out-of-plane and in-plane line-cut profiles of the blend films were shown in **Figure 5**. GIXD results revealed the molecular orientation and packing of the blend films before and after annealing treatment. The strong lamellar (100) diffraction peaks with q ≈ 0.26 Å$^{−1}$ in in-plane direction and the π-π (010) stacking peaks with q ≈ 1.7 Å$^{−1}$ in out-of-plane direction can be clearly observed for both PBT1-EH:ITCPTC and PBT1-EH:ITIC in as-cast blend films. The structural information came from PBT1-EH donor, which can be seen in pure film diffraction (**Figure 2d**). The weaker signals from ITIC and ITCPTC in blends were smeared by PBT1-EH polymer diffractions, and thus it is hard to analyze structure details in blends. The dominance of face-on orientation of both donor and acceptors in blend will facilitate carrier vertical transport between two electrodes. Thermal annealing can regulate the morphology of BHJ thin films. As can be seen from PBT1-EH:ITCPTC samples, annealing at 100 °C led to obvious improvement of PBT1-EH crystallization, since the signature peaks strongly improved (**Figure 5c and 5e**). PBT1-EH:ITIC blend showed less obvious structural changes under thermal treatment (**Figure 5b and 5d**), which is consistent with the fact that the surface morphology remained almost unchanged in AFM images (**Figure 4c and 4d**). It is interesting to note that thiophene modification of ITIC can affect the condensed-state structure in BHJ thin films. ITIC and PBT1-EH mix well in blends, and the existence of ITIC retards PBT1-EH crystallization. Thus thermal annealing does not affect device performance that much. In PBT1-EH:ITCPTC case, though a good molecular mixing exists in as spun thin film, thermal annealing leads to reorganization of PBT1-EH, which should lead to pronounced fibril growth when orientated in a face-on geometry. The demixing and polymer crystallization drive the morphology to a better optimum, boosting the device performance. The improved hole transport channels in annealed PBT1-EH:ITCPTC device lead to mobility enhancement. While a better electron transport of ITCPTC in amorphous state results in more balanced carrier extraction, making PBT1-EH:ITCPTC a better choice in device fabrication.

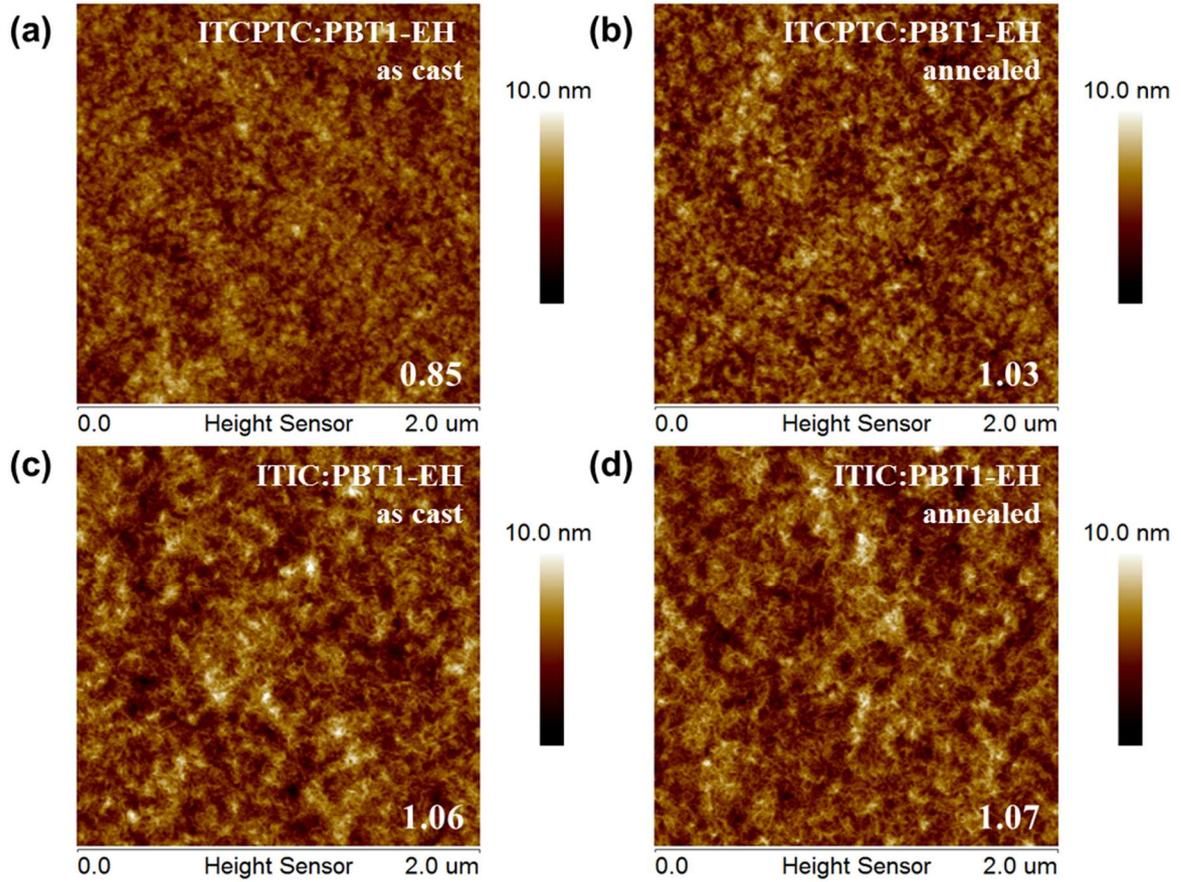

**Figure 4.** The AFM height sensor images: a) PBT1-EH:ITCPTC as-cast film; b) PBT1-EH:ITCPTC annealed film; c) PBT1-EH:ITIC as-cast film; d) PBT1-EH:ITIC annealed film.

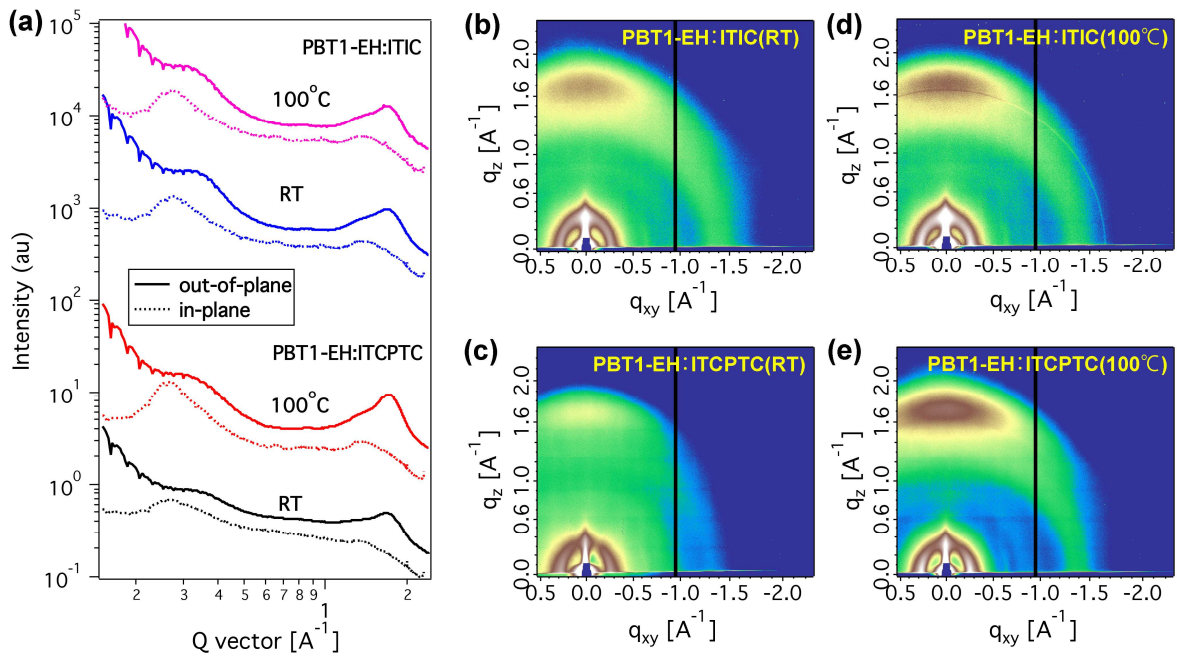

**Figure 5**. a) The out-of-plane and in-plane line-cut profiles of PBT1-EH:ITIC and PBT1-EH:ITCPTC blend films before and after annealing. The 2D GIXD patterns: b) PBT1-EH:ITIC as-cast film; c) PBT1-EH:ITCPTC as-cast film; d) PBT1-EH:ITIC annealed film; e) PBT1-EH:ITCPTC annealed film.

To show the microscopic morphology difference, and understand the dramatically enhanced FFs for the ITCPTC-based devices, we further investigated the phase separation of two blend films before and after thermal annealing by using resonant soft X-ray scatering (RSoXS). The corresponding scattering profiles were shown in **Figure 6**. The similar chemical nature of donor and acceptor materials makes the RSoXS characterization low in contrast when comparing to fullerene-based blends. X-ray photon energy 285.6 eV gives the best scattering signals. It is quite obvious that PBT1-EH:ITCPTC blends showed a much higher scattering intensity comparing to PBT1-EH:ITIC blends, and thus a much higher extent of phase separation existed in corresponding samples. The enhanced phase separation gives rise to improved fill factors as shown in previous report.[47,48] For PBT1-EH:ITIC blends, a characteristic length scale of phase separation was around 40 nm. These scattering profiles were quite similar, and thermal annealing did not change this size feature. Thus thermal annealing in this case does not strongly improve device performances. In contrast, PBT1-EH:ITCPTC blends showed a characteristic length scale of phase separation around 60 nm. Thermal annealing led to enhanced scattering intensity and reduction of phase separation size scale, both helping in improving device current. The enhanced scattering upon thermal annealing in PBT1-EH:ITCPTC blends suggests a phase purification process, which reveals the mechanism of morphology development when correlating with GIXD characterization:

PBT1-EH crystallization pushes ITCPTC molecules out and forms a better-defined bi-continuous fibril network with ITCPTC rich regions decorated in-between. Though slightly larger in phase separation length scale, generally good mixing in-between donor polymer and SMAs makes charge generation efficient. Better electron transport of ITCPTC molecule and larger crystal size made electron transport sufficient. Thus a reduced recombination was achieved. The simple modification of ITIC into ITCPTC thus not only change the material electronic structure, but also mediate the material interactions and crystallization, which in all are superior when combined with the promising high band gap PBT1-EH donor polymer.

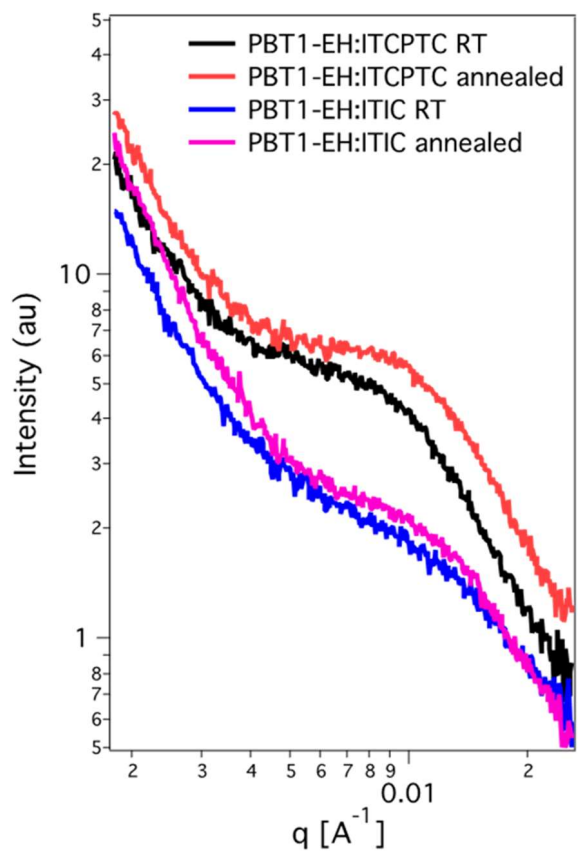

**Figure 6.** The RSoXS profiles of PBT1-EH:ITCPTC and PBT1-EH:ITIC in as-cast and annealing blend films.

In summary, we developed a novel small molecule acceptor of ITCPTC with thiophene-fused ending group by employing a new active methylene precursor of CPTCN. The ITCPTC based polymer solar cells with PBT1-EH as donor achieved very high PCEs of up to 11.8% with a remarkably enhanced fill factor (FF) of 0.751, a near 20% boost in PCE with respect to the ITIC based control device. To the best of our knowledge, these values are among the highest PCEs and FFs for the fullerene-free PSCs (Table S2). The BHJ active layer of PBT1-EH:ITCPTC exhibited a balanced carrier mobility ($\mu_e/\mu_h = 1.57$), a low charge recombination probability, strong π-π stacking in preferential face-on orientation and suitable phase separation, which contribute together to the excellent performance of ITCPTC based PSCs. We revealed that the simple modification of ITIC into ITCPTC not only change the material electronic structure, but also mediate the material interactions and crystallization. Taking these excellent advantages into consideration, we anticipate that the combination of the novel thiophene-fused ending group with other electron-rich block will create more promising SMAs. We are pretty sure that the new SMA of ITCPTC will set new records in fullerene-free PSCs. Developing more SMAs based on the new thiophene-fused ending group and mapping polymer donors for high-performance fullerene-free PSCs are under way.

**Acknowledgements**

D. Xie, T. Liu and W. Gao contributed equally to this work. This work was financially supported by the National Natural Science Foundation of China (NSFC) (No. 21572171 and 91433201), National Basic Research Program of China (973 Program 2013CB834805), the

[1]   H. Bin, L. Gao, Z. Zhang, Y. Yang, Y. Zhang, C. Zhang, S. Chen, L. Xue, C. Yang, M. Xiao, Y. Li, *Nat. Commun.* **2016**, *7*, 13651.

[2]   A. F. Eftaiha, J. Sun, I. G. Hill, G. C. Welch, *J. Mater. Chem. A* **2014**, *2*, 1201.

[3]   S. Li, L. Ye, W. Zhao, S. Zhang, S. Mukherjee, H. Ade, J. Hou, *Adv. Mater.* **2016**, *28*, 9423.

[4]   Y. Lin, X. Zhan, *Acc. Chem. Res.* **2016**, *49*, 175.

[5]   Y. Lin, F. Zhao, Y. Wu, K. Chen, Y. Xia, G. Li, S. K. K. Prasad, J. Zhu, L. Huo, H. Bin, Z. G. Zhang, X. Guo, M. Zhang, Y. Sun, F. Gao, Z. Wei, W. Ma, C. Wang, J. Hodgkiss, Z. Bo, O. Inganäs, Y. Li, X. Zhan, *Adv. Mater.* **2017**, *29*, 1604155.

[6]   C. B. Nielsen, S. Holliday, H.-Y. Chen, S. J. Cryer, I. McCulloch, *Acc. Chem. Res.* **2015**, *48*, 2803.

[7]   P. Sonar, J. P. Fong Lim, K. L. Chan, *Energy Environ. Sci.* **2011**, *4*, 1558.

[8]   Y. Yang, Z. Zhang, H. Bin, S. Chen, L. Gao, L. Xue, C. Yang, Y. Li, *J. Am. Chem. Soc.* **2016**, *138*, 15011.

[9]   W. Zhao, D. Qian, S. Zhang, S. Li, O. Inganäs, F. Gao, J. Hou, *Adv. Mater.* **2016**, *28*, 4734.

[10] Z. Zheng, O. M. Awartani, B. Gautam, D. Liu, Y. Qin, W. Li, A. Bataller, K. Gundogdu, H. Ade, J. Hou, *Adv. Mater.* **2016**, DOI: 10.1002/adma.201604241.

[11] Y. Lin, X. Zhan, *Mater. Horizons* **2014**, *1*, 470.

[12] H. Li, T. Earmme, G. Ren, A. Saeki, S. Yoshikawa, N. M. Murari, S. Subramaniyan, M. J. Crane, S. Seki, S. A. Jenekhe, *J. Am. Chem. Soc.* **2014**, *136*, 14589.

[13] H. Yao, Y. Chen, Y. Qin, R. Yu, Y. Cui, B. Yang, S. Li, K. Zhang, J. Hou, *Adv. Mater.* **2016**, *28*, 8283.

[14] D. Sun, D. Meng, Y. Cai, B. Fan, Y. Li, W. Jiang, L. Huo, Y. Sun, Z. Wang, *J. Am. Chem. Soc.* **2015**, *137*, 11156.

[15] X. Zhang, Z. Lu, L. Ye, C. Zhan, J. Hou, S. Zhang, B. Jiang, Y. Zhao, J. Huang, S. Zhang, Y. Liu, Q. Shi, Y. Liu, J. Yao, *Adv. Mater.* **2013**, *25*, 5791.

[16] Y. Lin, F. Zhao, Q. He, L. Huo, Y. Wu, T. C. Parker, W. Ma, Y. Sun, C. Wang, D. Zhu, A. J. Heeger, S. R. Marder, X. Zhan, *J. Am. Chem. Soc.* **2016**, *138*, 4955.

[17] Y. Li, D. Qian, L. Zhong, J. Lin, Z. Jiang, Z. Zhang, Z. Zhang, Y. Li, L. Liao, F. Zhang, *Nano Energy* **2016**, *27*, 430.

[18] Y. Lin, Y. Wang, J. Wang, J. Hou, Y. Li, D. Zhu, X. Zhan, *Adv. Mater.* **2014**, *26*, 5137.

[19] C. Zhang, T. Liu, W. Zeng, D. Xie, Z. Luo, Y. Sun, C. Yang, *Mater. Chem. Front.* **2017**, DOI:10.1039/c6qm00194g.

[20] Z. Luo, W. Xiong, T. Liu, W. Cheng, K. Wu, Y. Sun, C. Yang, *Org. Electron.* **2017**, *41*, 166.

[21] D. Meng, H. Fu, C. Xiao, X. Meng, T. Winands, W. Ma, W. Wei, B. Fan, L. Huo, N. L. Doltsinis, Y. Li, Y. Sun, Z. Wang, *J. Am. Chem. Soc.* **2016**, *138*, 10184.

[22] Q. Wu, D. Zhao, A. M. Schneider, W. Chen, L. Yu, *J. Am. Chem. Soc.* **2016**, *138*, 7248.

[23] T. Liu, D. Meng, Y. Cai, X. Sun, Y. Li, L. Huo, F. Liu, Z. Wang, T. P. Russell, Y. Sun, *Adv. Sci.* **2016**, *3*, 1600117.

[24] Y. Liu, C. Mu, K. Jiang, J. Zhao, Y. Li, L. Zhang, Z. Li, J. Y. L. Lai, H. Hu, T. Ma, R. Hu, D. Yu, X. Huang, B. Z. Tang, H. Yan, *Adv. Mater.* **2015**, *27*, 1015.

[25] P. E. Hartnett, A. Timalsina, H. S. S. R. Matte, N. Zhou, X. Guo, W. Zhao, A. Facchetti, R. P. H. Chang, M. C. Hersam, M. R. Wasielewski, T. J. Marks, *J. Am. Chem. Soc.* **2014**, *136*, 16345.

[26] T. Liu, Y. Guo, Y. Yi, L. Huo, X. Xue, X. Sun, H. Fu, W. Xiong, D. Meng, Z. Wang, F. Liu, T. P. Russell, Y. Sun, *Adv. Mater.* **2016**, *28*, 10008.

[27] O. K. Kwon, M. A. Uddin, J. H. Park, S. K. Park, T. L. Nguyen, H. Y. Woo, S. Y. Park, *Adv. Mater.* **2016**, *28*, 910.

[28] H. Li, Y. J. Hwang, B. A. E. Courtright, F. N. Eberle, S. Subramaniyan, S. A. Jenekhe, *Adv. Mater.* **2015**, *27*, 3266.

[29] X. Wu, W. Fu, Z. Xu, M. Shi, F. Liu, H. Chen, J. Wan, T. P. Russell, *Adv. Funct. Mater.* **2015**, *25*, 5954.

[30] S. Holliday, R. S. Ashraf, C. B. Nielsen, M. Kirkus, J. A. Röhr, C. Tan, E. Collado-Fregoso, A. C. Knall, J. R. Durrant, J. Nelson, I. McCulloch, *J. Am. Chem. Soc.* **2015**, *137*, 898.

[31] S. Holliday, R. S. Ashraf, A. Wadsworth, D. Baran, S. A. Yousaf, C. B. Nielsen, C. Tan, S. D. Dimitrov, Z. Shang, N. Gasparini, M. Alamoudi, F. Laquai, C. J. Brabec, A. Salleo, J. R. Durrant, I. McCulloch, *Nat. Commun.* **2016**, *7*, 11585.

[32] Y. Wu, H. Bai, Z. Wang, P. Cheng, S. Zhu, Y. Wang, W. Ma, X. Zhan, *Energy Environ. Sci.* **2015**, *8*, 3215.

[33] H. Bai, Y. Wang, P. Cheng, J. Wang, Y. Wu, J. Hou, X. Zhan, *J. Mater. Chem. A* **2015**, *3*, 1910.

[34] Y. Lin, Q. He, F. Zhao, L. Huo, J. Mai, X. Lu, C. Su, T. Li, J. Wang, J. Zhu, Y. Sun, C. Wang, X. Zhan, *J. Am. Chem. Soc.* **2016**, *138*, 2973.

[35] Y. Li, L. Zhong, F. Wu, Y. Yuan, H. Bin, Z. Jiang, Z. Zhang, Z. Zhang, Y. Li, L. Liao, *Energy Environ. Sci.* **2016**, *9*, 3429.

[36] Y. Li, X. Liu, F. Wu, Y. Zhou, Z. Jiang, B. Song, Y. Xia, Z. Zhang, F. Gao, O. Inganäs, Y. Li, L. Liao, *J. Mater. Chem. A* **2016**, *4*, 5890.

[37] Y. Lin, T. Li, F. Zhao, L. Han, Z. Wang, Y. Wu, Q. He, J. Wang, L. Huo, Y. Sun, C. Wang, W. Ma, X. Zhan, *Adv. Energy Mater.* **2016**, *6*, 1600854.

[38] Y. Lin, J. Wang, Z. Zhang, H. Bai, Y. Li, D. Zhu, X. Zhan, *Adv. Mater.* **2015**, *27*, 1170.

[39] W. Jiang, Y. Li, Z. Wang, *Chem. Soc. Rev.* **2013**, *42*, 6113.

[40] T. Liu, X. Pan, X. Meng, Y. Liu, D. Wei, W. Ma, L. Huo, X. Sun, T. H. Lee, M. Huang, H. Choi, J. Y. Kim, W. C. H. Choy, Y. Sun, *Adv. Mater.* **2016**, DOI 10.1002/adma.201604251.

[41] C. J. Brabec, A. Cravino, D. Meissner, N. Serdar Sariciftci, T. Fromherz, M. T. Rispens, L. Sanchez, J. C. Hummelen, *Adv. Funct. Mater.* **2001**, *11*, 374.

[42] L. Huo, T. Liu, X. Sun, Y. Cai, A. J. Heeger, Y. Sun, *Adv. Mater.* **2015**, *27*, 2938.

[43] Z. Tan, S. Li, F. Wang, D. Qian, J. Lin, J. Hou, Y. Li, *Sci. Rep.* **2014**, *4*, 4691.

[44] I. Riedel, J. Parisi, V. Dyakonov, L. Lutsen, D. Vanderzande, J. C. C. Hummelen, *Adv. Funct. Mater.* **2004**, *14*, 38.

[45] S. R. Cowan, A. Roy, A. J. Heeger, *Phys. Rev. B* **2010**, *82*, 245207.

[46] Y. Zang, C. Li, C. C. Chueh, S. T. Williams, W. Jiang, Z. Wang, J. Yu, A. K. Y. Jen, *Adv. Mater.* **2014**, *26*, 5708.

[47] L. Ye, S. Zhang, W. Ma, B. Fan, X. Guo, Y. Huang, H. Ade, J. Hou, *Adv. Mater.* **2012**, *24*, 6335.

[48] W. Ma, J. R. Tumbleston, M. Wang, E. Gann, F. Huang, H. Ade, *Adv. Energy Mater.* **2013**, *3*, 864.

**A novel small molecule acceptor of ITCPTC with thiophene-fused ending group is designed and synthesized.** The ITCPTC based polymer solar cells with PBT1-EH as donor achieved very high PCEs of up to 11.8% with a remarkably enhanced fill factor of 0.751, a near 20% boost in PCE with respect to the ITIC based control device.

keywords: small molecular acceptor, ending group, fullerene-free polymer solar cell, efficiency

Dongjun Xie[†], Tao Liu[†], Wei Gao[†], Cheng Zhong, Lijun Huo, Zhenghui Luo, Kailong Wu, Wentao Xiong, Feng Liu,* Yanming Sun* and Chuluo Yang*

**A Novel Thiophene-fused Ending Group Enabling an Excellent Small Molecule Acceptor for High-performance Fullerene-free Polymer Solar Cells with 11.8% Efficiency**

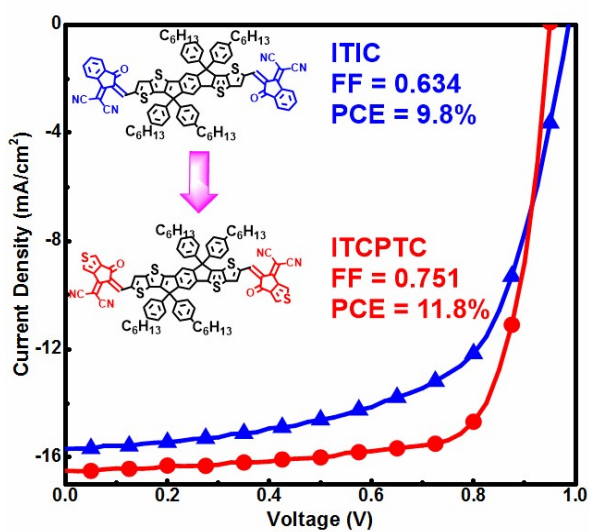